# Intuitive Proof of Black-Scholes Formula Based on Arbitrage and Properties of Lognormal Distribution

Alexei Krouglov

*796 Caboto Trail, Markham, Ontario L3R 4X1, Canada*

*Abstract*

Presented is intuitive proof of Black-Scholes formula for European call options, which is based on arbitrage and properties of lognormal distribution. Paper can help students and non-mathematicians to better understand economic concepts behind one of the biggest achievements in modern financial theory.

*Introduction*

Traditional derivation of Black-Scholes formula [1] requires employment of stochastic differential equations and Ito calculus. It makes this subject pretty challenging for students and people not fluent in those advanced mathematical subjects. Current article shows deduction of Black-Scholes formula based purely on the concept of arbitrage and properties of lognormal distributions.

*Concept of Arbitrage*

Concept of arbitrage says that when the future price of investment asset is unknown it is assumed that the price of asset today with its delivery in the future is determined by some other asset whose future price is deterministic.

Let me show how concept of arbitrage works with the price of non-dividend-paying stock share. Assume that at time $t = t_0$ the price of share is $S_0 = S(t_0)$. We want to determine the price of share at time $t = t_0$ with delivery of share at time $t = t_0 + T$. I denote that price as $F_0^T$. We don't know how the price of share $S(t)$ will change in time interval $t_0 \leq t \leq t_0 + T$.

Now I assume that we have another investment asset whose price is deterministic in the future. That asset is a bank account that pays continuous interest rate $r$. Thus if one puts in bank at time $t = t_0$ the amount of cash equal to $S_0$ she can take back from bank at time $t = t_0 + T$ the amount of cash equal to $S_0 e^{rT}$. Likewise if one borrows from bank at time $t = t_0$ the amount of cash equal to $S_0$ she have to give back to bank at time $t = t_0 + T$ the amount of cash equal to $S_0 e^{rT}$. And exactly same amount determines the market price of share at time $t = t_0$ with delivery of share at time $t = t_0 + T$ i.e. $F_0^T = S_0 e^{rT}$.

Why? Because if $F_0^T > S_0 e^{rT}$ market participants at time $t = t_0$ can borrow money from bank to buy $N$ shares for amount $N S_0$ and sell these shares right away with delivery at time $t = t_0 + T$ for amount $N F_0^T$. Then market participants deliver $N$ shares for a payment in $N F_0^T$ and pay back amount $N S_0 e^{rT}$ to the bank at time $t = t_0 + T$. Thus the profit is equal to $N(F_0^T - S_0 e^{rT}) > 0$. This profit opportunity is exposed to all market participants, therefore demand on shares drives their price higher than today's price equal to $S_0$ and supply of shares with delivery at time $t = t_0 + T$ drives their price lower on the market than today's price equal to $F_0^T$ until eventually $F_0^T = S_0 e^{rT}$.

Also if $F_0^T < S_0 e^{rT}$ market participants who own $N$ shares at time $t = t_0$ sell their shares for amount $N S_0$, put money to bank and buy shares back with delivery at time $t = t_0 + T$ for amount $N F_0^T$. At time $t = t_0 + T$ market participants withdraw amount $N S_0 e^{rT}$ from the bank and receive again $N$ shares for a payment in $N F_0^T$. Therefore profit of market participants is equal to $N(S_0 e^{rT} - F_0^T) > 0$ with no penalty. That profit opportunity will definitely be exploited. Therefore supply of shares drives their price lower than today's price equal to $S_0$ and demand on shares with delivery at time $t = t_0 + T$ drives their price higher than today's price equal to $F_0^T$ until eventually $F_0^T = S_0 e^{rT}$.

*Probabilistic Nature of Options*

Stock options represent the right of investors to buy the shares of stock in the future at predetermined price. Thus for example the price of European call options today depicts the future prices of stock shares multiplied by probability that the price of shares exceeds this predetermined price (so-called strike price). And that probability obviously depends on the stock volatility. Therefore if stock is more volatile it gives one bigger profitable opportunities to exercise her right to buy stock shares at the predetermined price.

But we are unaware of the future prices of stock shares, aren't we? Yes, but we employ the concept of arbitrage, which says since the future prices of stock shares are unknown market assumes that their prices today are determined by another asset with deterministic future prices as for example the bank account paying a continuous interest rate $r$.

So the problem is only with choice of assumptions regarding to the volatility of stock shares. Academics have studied stock prices observed by the market and decided that their probabilistic behavior is well approximated by the lognormal distribution [2] (i.e. logarithm of stock prices is approximately normally distributed).

*Some Properties of Lognormal Distribution*

Here I discuss some properties of lognormal distributions [3], which will be used later.

Random variable $x_L$ that is continuously distributed in interval $0 < x_L < +\infty$ is said to have lognormal distribution described by probability density function $f_L(x_L)$ if variable $x_N$, that is defined as $x_N = \ln x_L$, has normal distribution described by probability density function $f_N(x_N)$ in interval $-\infty < x_N < +\infty$. I denote the mean and variance of normally distributed variable $x_N$ as $\mu_N$ and $\sigma_N^2$ respectively. Then the mean $\mu_L$ of variable $x_L$ with lognormal distribution, which is defined as

$$\mu_L = \int_0^{+\infty} x_L f_L(x_L) dx_L, \qquad (1)$$

is calculated according to following formula,

$$\mu_L = e^{\left(\mu_N + \frac{1}{2}\sigma_N^2\right)}. \qquad (2)$$

I introduce truncated or partial distribution of variable $x_L$ in interval $k < x_L < +\infty$ where $k > 0$. For this partial distribution of variable $x_L$ one can consider partial expectation $\mu_L(k)$ defined as

$$\mu_L(k) = \int_k^{+\infty} (x_L - k) f_L(x_L) dx_L \qquad (3)$$

and calculate $\mu_L(k)$ as

$$\mu_L(k) = e^{\left(\mu_N + \frac{1}{2}\sigma_N^2\right)} \Phi\left(\frac{-\ln k + \mu_N + \sigma_N^2}{\sigma_N}\right) - k \Phi\left(\frac{-\ln k + \mu_N}{\sigma_N}\right) \qquad (4)$$

where $\Phi(y)$ is cumulative distribution function for standard normally distributed random variable $y_N$ with probability density function $f_N(y_N)$ i.e. normally distributed variable $y_N$ has mean $\mu_N(y_N) = 0$ and variance $\sigma_N^2(y_N) = 1$; function $\Phi(y)$ defined as following

$$\Phi(y) = \int_{-\infty}^{y} f_N(y_N) dy_N. \qquad (5)$$

Now we are ready to calculate Black-Scholes formula for European call options.

### *Justification of Black-Scholes Formula*

Here I obtain the formula to calculate the price of European option on a non-dividend paying stock.

Let me assume that at time $t = t_0$ the price of share is $S_0 = S(t_0)$. I want to find the price of option at time $t = t_0$ to be executed at time $t = t_0 + T$ with strike price $k$. I denote the price of option as $C_0^T(k)$.

As before, market doesn't know how the price of share $S(t)$ will vary in the time interval $t_0 \leq t \leq t_0 + T$ but according to the arbitrage principle market has to assume that price of share at time $t = t_0$ with delivery of share at time $t = t_0 + T$ is $F_0^T = S_0 e^{rT}$. Thus at time $t = t_0$ the price of share equal $S_0 e^{rT}$ is the expected market price at time $t = t_0 + T$.

I mentioned above that academics believe that stock prices can be approximated by the lognormal distribution. Thus at time $t = t_0$ the expected value $\mu_L$ of share's price $F_0^T$ at time $t = t_0 + T$ is equal to $\mu_L = S_0 e^{rT}$.

What about stock volatility? Academics assume that market participants at time $t = t_0$ imply that volatility of stock prices at time $t = t_0 + T$ should depend on the length of time interval $T$. Actually people prefer to express the volatility of lognormally distributed stock in terms of a matching normal distribution $x_N$ and suppose that the variance of logarithm of stock prices is equal to $\sigma_N^2 = \sigma^2 T$.

Likewise one can express the expected value $\mu_L$ of share's price $F_0^T$ in terms of the expected value $\mu_N$ for the corresponding normal distribution $x_N$,

$$\mu_L = e^{\left(\mu_N + \frac{1}{2}\sigma_N^2\right)} = e^{\left(\mu_N + \frac{1}{2}\sigma^2 T\right)} = S_0 e^{rT}.$$

Therefore it takes place,

$$\mu_N = \ln S_0 + \left(r - \frac{1}{2}\sigma^2\right)T. \tag{6}$$

Eventually the price of European call option $C_0^T(k)$ can be evaluated using the expression for partial expectation $\mu_L(k)$,

$$\mu_L(k) = e^{(\ln S_0 + rT)} \Phi\left(\frac{-\ln k + \ln S_0 + rT + \frac{1}{2}\sigma^2 T}{\sqrt{\sigma T}}\right) - k\Phi\left(\frac{-\ln k + \ln S_0 + rT - \frac{1}{2}\sigma^2 T}{\sqrt{\sigma T}}\right).$$

Since $\mu_L(k)$ represents what market expects at time $t = t_0$ how the price of European call option $C_0^T(k)$ becomes worth at time $t = t_0 + T$, then we have an equality $\mu_L = C_0^T(k)e^{rT}$ according to the arbitrage principle or equivalently

$$C_0^T(k)e^{rT} = S_0 e^{rT} \Phi\left(\frac{\ln\frac{S_0}{k} + \left(r + \frac{1}{2}\sigma^2\right)T}{\sqrt{\sigma T}}\right) - k\Phi\left(\frac{\ln\frac{S_0}{k} + \left(r - \frac{1}{2}\sigma^2\right)T}{\sqrt{\sigma T}}\right)$$

and ultimately [4],

$$C_0^T(k) = S_0 \Phi\left(\frac{\ln\frac{S_0}{k} + \left(r + \frac{1}{2}\sigma^2\right)T}{\sqrt{\sigma T}}\right) - k e^{-rT} \Phi\left(\frac{\ln\frac{S_0}{k} + \left(r - \frac{1}{2}\sigma^2\right)T}{\sqrt{\sigma T}}\right). \tag{7}$$

*Summary*

Article demonstrates that in order to get Black-Scholes formula for calculation of the price for European call options it is sufficient for one to use logic based on a number of assumptions where some assumptions have a firm economic nature and another ones are purely probabilistic assumptions.

Principal economic assumption says when future prices of investment asset are unknown market assumes that the price of asset today with its delivery in the future is determined by another asset such as a bank account whose future prices are deterministic.

Probabilistic assumptions say when future prices of such investment asset as shares of non-dividend-paying stock are unknown market assumes that the prices of stock shares at the end of investment period will have lognormal distribution and that volatility of the prices of stock shares can be explained by the variance of logarithm of prices that is directly proportional to the length of investment period.

Logic besides assembling assumptions above into Black-Scholes formula is as following. Market assumes today that the price of stock shares in the future is determined by the growth of cash investments in bank accounts. Stock option is a right of investor to buy the shares of stock in the future at predetermined price. Thus today market anticipates that at the end of investment period monetary value of European call options will become equal to the expected positive cash surplus between possible prices of stock shares and their predetermined price. The expected surplus between possible prices of stock shares and the predetermined price of stock shares is calculated based on the volatility of prices of stock shares around the expected price of stock shares at the end of investment period. In turn the expected surplus between prices of stock shares and the predetermined price of them reveals how market anticipates today cash value of European call option at the end of investment period. And to obtain today's price of European call option on the market from cash value of European call option at the end of investment period (since exact value growth of the investment asset like European call option is unknown) market uses the same economic mechanism as it employs in bringing the future values of cash investments in bank accounts to their today's value i.e. in other words market discounts value of European call option as future cash value in bank accounts to today's cash value.

Thus discount of the expected positive surplus between possible prices of stock shares and their predetermined price with help of continuous interest rate over the length of investment period finally produces the Black-Scholes formula for European call options.